\documentclass[]{jfm}
\usepackage{graphicx}
\usepackage{newtxtext}
\usepackage{newtxmath}
\usepackage{natbib}
\usepackage{hyperref}
\usepackage{subfigure}
\usepackage{float}
\usepackage{verbatim}
\usepackage{color}
\usepackage{ulem}

\title{The effect of collision-coagulation on the mean relative velocity of particles in turbulent flow: systematic results and validation of model.}

\author{Xiaohui Meng\aff{1},
	\and Ewe-Wei Saw\aff{2}
	\corresp{\email{ewsaw3@gmail.com}}}

\affiliation{\aff{1}School of Atmospheric Sciences, Sun Yat-Sen University, Zhuhai, China
	\aff{2}School of Atmospheric Sciences and Guangdong Province Key Laboratory for Climate Change and Natural Disaster Studies, Sun Yat-Sen University, Zhuhai, China}

\newcommand{\ew}[1]{\textcolor{black}{#1}}
\newcommand{\xh}[1]{\textcolor{black}{#1}}

\begin{document}
	\maketitle

\begin{abstract}
	
	\ew{The mean radial component of relative velocity (MRV) between pairs of inertial particles is studied, where the particles are advected by turbulent flow and undergo collision-and-coagulation. A previously proposed phenomenological model of MRV for low-inertia particles \citep{saw2022intricate} is corrected (improved) and shown to produce better predictions of the MRV as a function of particle separation distance $r$. Then, using direct numerical simulation (DNS), the relationship between the DNS-produced MRV and particle/turbulent parameters is studied. The results show that for particles with near-zero Stokes numbers ($St$), the MRV is roughly independent of $St$. At larger $St$, the magnitude of MRV increases with $St$ and this change is most pronounced when $St>0.2$. Assuming that the relative particle velocities are derived from fluid velocity differences associated with a nominal resonant length scale, we find an empirical relation between this resonant scale and Stokes number such that the resonant scale has the form: $d + \alpha St^{\beta}$, where $\beta \approx 1.86$. We show that, coupled with this empirical result, the aforementioned MRV model could be extended to predict MRV for any finite $St$, and we further show that the predictions are accurate against the DNS results. Our results also suggest that the extended model could also accurately account for possible Reynolds number ($Re_{\lambda}$) effect by simply allowing $\alpha$ and $\beta$ to be functions of $Re_{\lambda}$. We also find that when the particle diameter is smaller than the Kolmogorov length scale, the MRV for particles with the same $St$ is independent of the particle diameter. By studying the MRV under different Reynolds numbers ($Re_{\lambda} = 84, 124, 189$), we find that for particles with $St\ll1$, the MRV is independent of $Re_{\lambda}$. For larger $St$, $Re_{\lambda}$ dependence is observed such that the coefficients $\alpha$ and $\beta$ decrease with $Re_{\lambda}$.} 

\end{abstract}

\section{Introduction}

The collision and coagulation of inertial particles in turbulent flow is a common and important physical process in both scientific research and industrial applications. For example, the collision rate between small droplets in atmospheric clouds affects precipitation \citep{shaw2003particle,grabowski2013growth}. In astrophysical environments such as protoplanetary disks, the collision of dust grains is a step in the formation of planetesimal \citep{johansen2009particle,pan2011turbulent}. In turbulent spray combustion, the distribution and the size of fuel droplets affect the efficiency of energy conversion \citep{smith2002significance}. \citep{saffman1956collision} studied the collision rate of no-inertia particles in turbulent flow and obtained the collision rate: $N_c=n_1n_2(r_1+r_2)^3(\frac{8\pi\varepsilon}{15\nu})^{1/2}$, in which $n_1$ and $n_2$ are the mean concentrations of two kinds of particles, respectively, $r_1$ and $r_2$ are their radii. The mean radial relative velocity of zero-inertia particles is $(r_1+r_2)^3(\frac{8\pi\varepsilon}{15\nu})^{1/2}$. Unlike fluid particles, inertial particles are preferentially concentrated in the high strain and low vorticity region of turbulent flow, resulting in an inhomogeneous distribution, which is also called particle clustering \citep{squires1991preferential,saw2008inertial}. In addition, the relative velocity of inertial particles increases at contact in turbulent flow \citep{bragg2014new}. These two effects make the collision rate of inertial particles in turbulent flow is larger than that of fluid particles and difficult to calculate. In \citep{sundaram1997collision}, the relationship among the collision rate, particle clustering and particle relative radial velocity was presented: $N_c/(n_1n_2V)=4\pi d^2g(d)\langle w_r(d)\rangle$, in which $d$ is the particle diameter, $g(d)$ is the radial distribution function (RDF) when the particle separation distance $r$ is equal to $d$, and $\langle w_r(d)\rangle$ is the mean radial relative velocity (MRV) of particles. $V$ is the spatial volume of the domain. This finding makes it possible to study the particle RDF and MRV independently and to predict the collision rate of inertial particles in different situations. This paper focuses on the MRV of inertial particles, and the following also focuses on the current state of research related to the relative velocity of particles. 

The particle Stokes number is a dimensionless quality characterizing the inertia of particles, is equal to the ratio of the particle relaxation time $\tau_p$ to the Kolmogorov time scale $\tau_{\eta}$ of the flow. The theoretical studies show that for particles with medium Stokes number, they have a path-history effect in turbulent flow. When particles are at the same position, their velocities have low correlation, resulting in high relative velocity. This phenomenon is called sling effect or caustics \citep{wilkinson2005caustics,wilkinson2006caustic,falkovich2002acceleration}. The mean radial relative velocity is the relative velocity of particles projected in the radial direction: $\langle w_r\rangle=\langle(\vec{v}_2-\vec{v}_1)\cdot\vec{r}\rangle$, which is also the first-order structure function of particle relative velocity. If the collision-coagulation of particles is not considered, since the turbulent velocity is isotropic at small scale, $\langle w_r\rangle=0$. In most numerical simulation studies, collision and coagulation of particles are not considered, therefore these works often investigated higher order particle relative velocity structure function, the second-order structure function $S_{\parallel}^2$ for example, or particles mean negative radial relative velocity $\langle w_r\vert_{w_r<0}\rangle$, to indirectly obtain the trend of the MRV. The results show that the value of $S_{\parallel}^2$ and $\langle w_r\vert_{w_r<0}\rangle$ of the particles increase gradually with the increase of the particle Stokes number, and the growth is significant when $St>0.2$, which is consistent with the finding of the sling effect/caustics \citep{bewley2013observation,ireland2016effect}. In \citep{falkovich2002acceleration} and \citep{wilkinson2005caustics}, they think that the sling effect or caustics becomes more significant with the increase of the turbulent Reynolds number. However, the numerical results show that for small $St$ particles, $S_{\parallel}^2$ and $\langle w_r\vert_{w_r<0}\rangle$ are only weakly dependent on turbulent Reynolds number \citep{ireland2016effect}. For particles with $St>3.0$, they decrease with increasing Reynolds number \citep{bec2010intermittency,rosa2013kinematic}. 

In \citep{saw2022intricate}, they find that when particles collide and coagulate, the magnitude of MRV increases significantly in the range of the separation distance $r$ is close to the particle diameter $d$. They also propose a MRV phenomenological model to illustrate the relationship between MRV and the separation distance $r$. Based on their results, we first make an improvement for the MRV phenomenological model. Then, the collision-coagulation is considered in our numerical simulation, the relationship between the MRV with particles and turbulent parameters is studied directly. 
The contents of this paper are organized as follows. In Section 2, we describe the numerical method that are used to simulate homogeneous isotropic turbulence and particle motion with the consideration of particle collision-coagulation. The improvement of the MRV phenomenological model is shown in section 3. The results from DNS are discussed in Section 4. A summary and main conclusion are given in Section 5.

\section{Simulation method}

Direct numerical simulation (DNS), which fully solves the Navier-Stokes equation in both spatial and temporal space, is used to study the particle laden flow in this paper. The incompressible Navier-Stokes equations are solved in a cube which has periodic boundary condition in three directions and the length of which is $2\pi$:

\begin{equation}
	\frac{\partial\vec{u}}{\partial t}+\vec{u}\cdot\nabla\vec{u}=-\frac{1}{\rho}\nabla p+\nu\nabla^{2}\vec{u}+\vec{f}(\vec{x},t)
	\label{NS}
\end{equation}

\begin{equation}
	\nabla\cdot\vec{u}=0
	\label{incompressible}
\end{equation}

Where $\vec{u}$, $p$, $\rho$ and $\nu$ are velocity, pressure, density and the kinetic viscosity of turbulent flow. $\vec{f}$ is the forcing scheme, added in the large scale region of turbulence to maintain the statistics stationary \citep{eswaran1988examination}. The N-S equations are transformed from physical space to wavenumber space, which is called the pseudo-spectral method. The 2/3-method is used to deal with aliasing error raised by the convection term in \ref{NS} \citep{rogallo1981numerical}.

The influence of the turbulent Reynolds number on the particle MRV is studied in this paper, the range of Reynolds number we choose is $Re_{\lambda}=84\sim189$, where $Re_{\lambda}$ is the Taylor microscale Reynolds number. In different cases, $k_{max}\eta=1.59$, 1.21, 1.38, where $k_{max}$ is the maximum wavenumber, and Courant number $C=0.0248$, 0.0401, and 0.0865 respectively. The detailed parameters are shown in table \ref{tab1}. The simulation method and the DNS parameters are the same with \citep{meng2023sharp}. In \citep{meng2023sharp}, a higher resolution simulation is conducted to study the possible effects of the sub-Kolmogorov intermittency on the RDF of particle, using $N=1024$ at $Re_{\lambda}=124$. The result indicates that the effect of unresolved sub-Kolmogorov intermittency may modify slightly the inertial clustering exponent but would not cause any significant changes to the qualitative trends of particle RDF. Therefore, it can be considered that the sub-Kolmogorov intermittency will not have effect on the main results in this paper.

\begin{table}
	\begin{center}
		\begin{tabular}{c|cccccccccc}
			& $N$ & $\nu$ & $\varepsilon$ & $u^{\prime}$ & $\lambda$ & $\eta$ & $\tau_{\eta}$ & $L$ & $T_{L}$ & $Re_{\lambda}$ \\
			flow 1 & 256 & 0.001 & 0.0326 & 0.3519 & 0.2386 & 0.0132 & 0.1750 & 0.5073 & 1.4416 & 84 \\
			flow 2 & 256 & 0.001 &  0.1013 & 0.5684 & 0.2187 & 0.0100 & 0.0993 & 0.6151 & 1.0822 & 124 \\
			flow 3 & 512 & 0.001 &  0.9472 & 1.226 & 0.1544 & 0.0057 & 0.0325 & 0.7398 & 0.6034 & 189 \\
		\end{tabular}
		\caption{The DNS parameters and time-averaged statistics. $N$ is the simulation grid size, $\nu$ is the kinematic viscosity of turbulence, $\varepsilon$ is the dissipation rate of turbulent flow, $u^{\prime}$ is the root-mean-square velocity of turbulent flow, $\lambda$ is the Taylor length scale, $\eta$ and $\tau_{\eta}$ are the Kolmogorov length and time scale, $L$ and $T_{L}$ are the integral length and time scale, $Re_{\lambda}$ is the Taylor scaled Reynolds number.}
	\label{tab1}
	\end{center}
\end{table}

Particles in this paper are small and heavy, the particle diameter $d$ is much smaller than the Kolmogorov length scale $\eta$ and the particle density $\rho_p$ is much larger than the density of flow $\rho$. The basic physical process of particle collision-coagulation is considered in this paper, therefore the gravitational effect and the inter-particle hydrodynamic interaction are not included in simulation system. Particles and turbulent flow are only one-way coupling. Under these hypotheses, particles are only conducted by the Stokes drag force, the motion equation of particles is shown below \citep{maxey1983equation}:

\begin{equation}
	\frac{d\vec{v}}{dt}=\frac{\vec{u}(\vec{x})-\vec{v}}{\tau_{p}}
\end{equation}
where $\vec{v}$ is particle velocity, $\vec{u}(\vec{x})$ is the flow velocity at particle position, and $\tau_p$ is particle relaxation time.

Particles will collide and create a new particle with the conservation of momentum and mass when the particle separation distance $r$ is equal to or less than the sum of the particle radii. To avoid successive collisions with larger particles created by collision-coagulation, larger particles are removed from the system as soon as they are created. In this situation, in order to maintain the equilibrium of the particles in the system, a fixed number of particles are added to the system at certain time steps, and these new particles have the same parameters as the initial particles. After 10 turbulent characteristic integration times, the particles in the system reach stability. It is important to note that, only monodisperse particles are considered in the calculation of the MRV, which means the relative velocity between particles with different $St$ is not taken into account. The influences of particle Stokes number and the particle size on MRV are studied in this paper, the range of $St$ we choose is $St=0.01~2.0$ and the particle diameters are $d=1/3d_*$, $d_*$, and $3d_*$, where $d_*=9.49\times10^{-4}$. It should be noted that the diameter size of the particles is only used to determine whether collisions occur between the particles, the inter-particle hydrodynamic is not considered. 

\section{Correction (improvement) of the MRV phenomenological model}\label{sec.Model}

\ew{Here, we present a correction (improvement) of the MRV phenomenological model proposed by \cite{saw2022intricate} and show that this substantially improve the model's accuracy. We begin with a brief introduction of the model. The model describes the relationship between MRV and the particle separation distance $r$ in the $St \ll 1$ limit, such that:}

\begin{equation}
	\begin{split}
	\langle w_{r}\rangle & =p_{-}\langle w_{r}\vert w_{r}<0\rangle+p_{+}\langle w_{r}\vert w_{r}\geq0\rangle \\
	&\approx -p_{-}\,\xi_{-}\,r \,+ \, p_{+}\,\xi_{+}\,r \left[1+\frac{\int_{\theta_m}^{0}P_{\theta}^{+}(\theta^{\prime})cos(\theta^{\prime})d\theta}{\int_{0}^{\pi/2}P_{\theta}^{+}(\theta^{\prime})cos(\theta^{\prime})d\theta^{\prime}} \right]
	\end{split}
	\label{Eq.wr_fit}
\end{equation}
\ew{where $p_+$ ($p_-$) is the probability for a sample of $w_r$ to turn out positive (negative) and the constants $\xi_{+}$, $\xi_{-}$ depends on the flow energy dissipation rate \citep{saw2022intricate}. As a first-order account, the skewness in the distribution of particle relative velocities were neglected which leads to $p_+\equiv\int_0^{\frac{\pi}{2}}P_{\theta}d\theta=0.5$, and $p_- \equiv 1 - p_+ =0.5$. $P_{\theta}^+$ is a conditional probability density function (PDF) of $\theta$ defined as $P_{\theta}^+\equiv P(\theta \,\, \vert \, w_r\leq0)\equiv P(\theta \,\, \vert \, \theta\in\left[0,\pi/2\right])$. $\theta$ is the angle between the relative velocity $\vec{w}$ and the relative displacement $\vec{r}$ of a particle-pair. \cite{saw2022intricate} argue that PDF of $\theta$ may be constructed using the (statistical) central-limit theorem, such that:}

\begin{equation}
P(\theta)=Nexp\left[Kcos(\theta-\mu_{\theta})\right]sin(\theta)
\label{Eq.Ptheta}
\end{equation}

\ew{where $Nexp[…]$ is the circular normal distribution, and $K$, a variance parameter, may be inferred by matching the model produced transverse-to-longitudinal ratio of structure functions (TLR) of the particle relative velocities to the one measured directly in experiment or DNS \citep{saw2022intricate}. In \citep{saw2022intricate}, the model predicts MRV that agrees best with DNS outcome when $K$ is calibrated using the fourth-order structure functions instead of the second-order which one would normally expect since MRV is a low order statistics. This suggests that the model may be incomplete\citep{saw2022intricate}.}

\begin{figure}
\centerline{
\subfigure[]{
\includegraphics[width=0.49\textwidth]{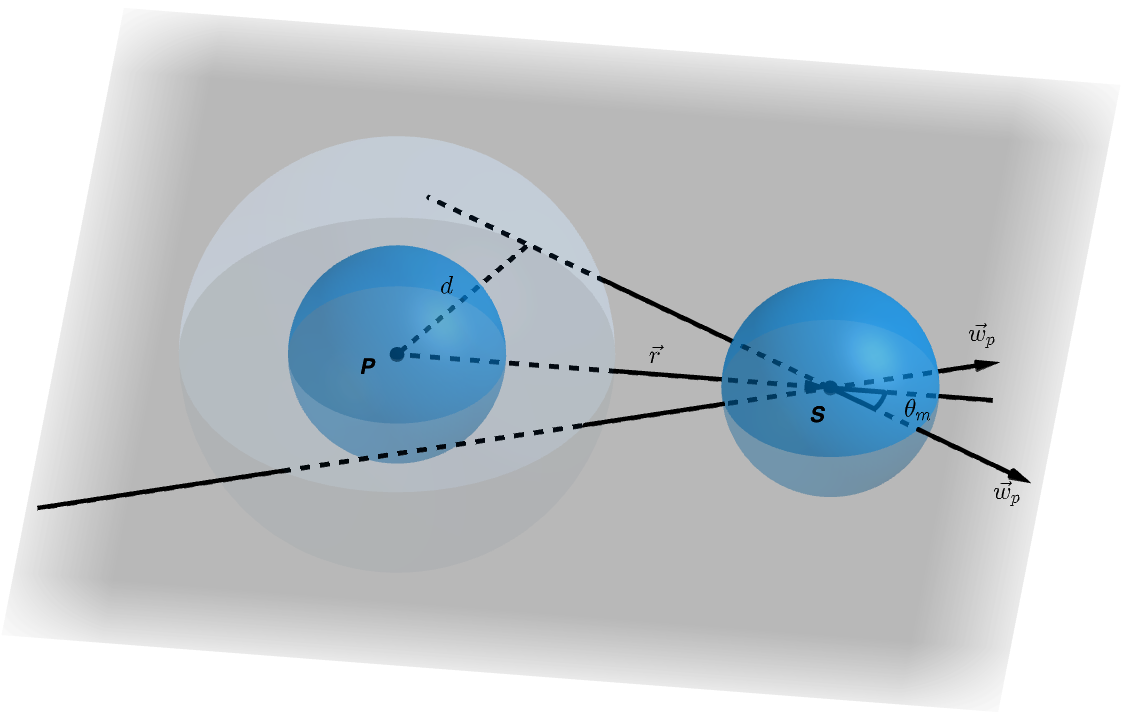}
\label{Fig.pcollision}
}
\subfigure[]{
\includegraphics[width=0.41\textwidth]{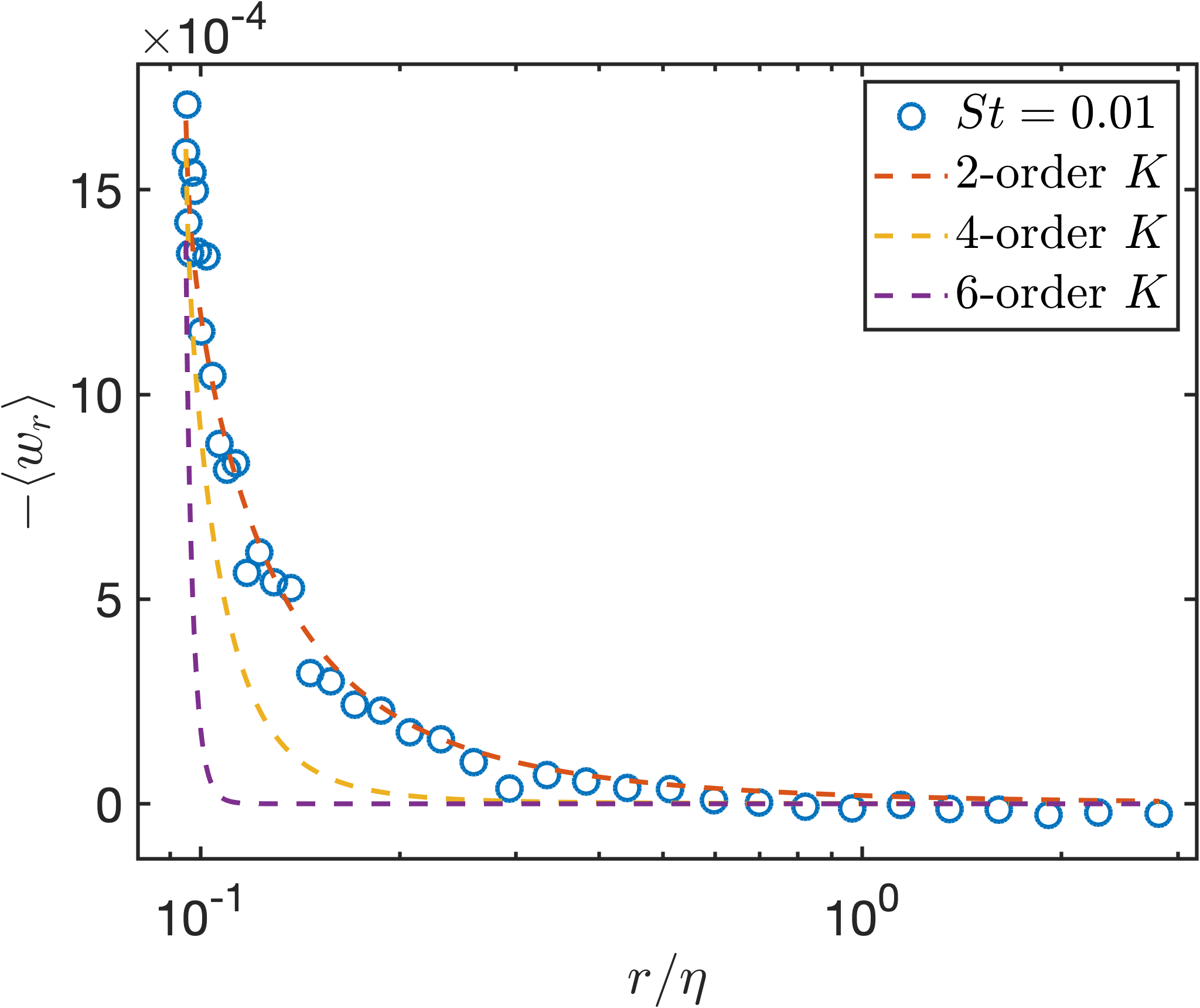}
}
\label{Fig.mrv_compare}}
\caption{(a). Schematic illustrating the ideas behind the MRV phenomenological model in \citep{saw2022intricate} that explain the change of MRV in the regime of $r\approx d$. $\vec{r}$ is the relative position of the S particle to the P particle, while $\vec{w}$ represents the relative velocity of S to P. (b). The MRV for particles with $St=0.01$ compared with prediction via the phenomenological model (Eq.\ref{Eq.wr_fit}) using new $p_+(p_-)$ calculation method. DNS result: $\circ$: $St=0.01$. Dashed lines are model predictions using Eq.\ref{Eq.wr_fit} and Eq.\ref{Eq.Ptheta} with K obtained by a certain order of structure function of particle relative velocity. red: second-order, yellow: fourth-order, purple: sixth-orther.}
\label{Fig.mrv_model}
\end{figure}

\ew{According to the idealized picture of particle collision-coagulation process described in \citep{saw2022intricate}, as shown in Figure \ref{Fig.pcollision}, the small scale trajectory of particle can be regarded as rectilinear at separation $r$ close to the particle diameter $d$. Since particles collide and coagulate when $r= d$, according to the geometrical analysis, the angle between $\vec{w}$ and $\vec{r}$ cannot be smaller than $\theta_m$, which is a value related to the ratio of $d$ to $r$: $\theta_m=sin^{-1}(d/r)$. This lower bound of $\theta$ affects the symmetry of the PDF of particle relative velocity, therefore $p_+$ (also $p_-$) should not be fixed at 0.5 but should equals a value that varies with $\theta_m$ (which in turns varies with $r$). Thus, we change the probability of a realization of $w_r$ being positive and negative to: 
}

\begin{equation}
p_+=\frac{\int_{\theta_m}^{\pi/2}P_{\theta}d\theta}{\int_{\theta_m}^{\pi}P_{\theta}d\theta}, \,\, p_-=\frac{\int_{\pi/2}^{\pi}P_{\theta}d\theta}{\int_{\theta_m}^{\pi}P_{\theta}d\theta},
\label{Eq.pp}
\end{equation}
\ew{which also satisfy $p_+ +p_-=1$. 
} 

\ew{Bringing the revised $(p_+, p_-)$  into Eq.\ref{Eq.wr_fit} and using Eq.\ref{Eq.Ptheta} where $K$ is calibrated by the measured (via DNS) second-, fourth-, and sixth-order structure functions of particle relative velocity respectively, the new MRV model predictions is compared with the DNS results, as shown in  Figure \ref{Fig.mrv_model}. We see that the second-order calibration produces result with the best agreement with DNS, and which is significantly better than that in \citep{saw2022intricate}. This provide strong support for the validity of the MRV model since second-order statistics are the most appropriate (orthodox) reference for the determination of $K$, which is itself a variance parameter analogous to $1/\sigma^2$ in the Gaussian distribution.}


\section{Results and Discussion}\label{sec.Result}

\subsection{Stokes number dependence}\label{sec.St}

\begin{figure}
\centerline{
\includegraphics[width=0.95\textwidth]{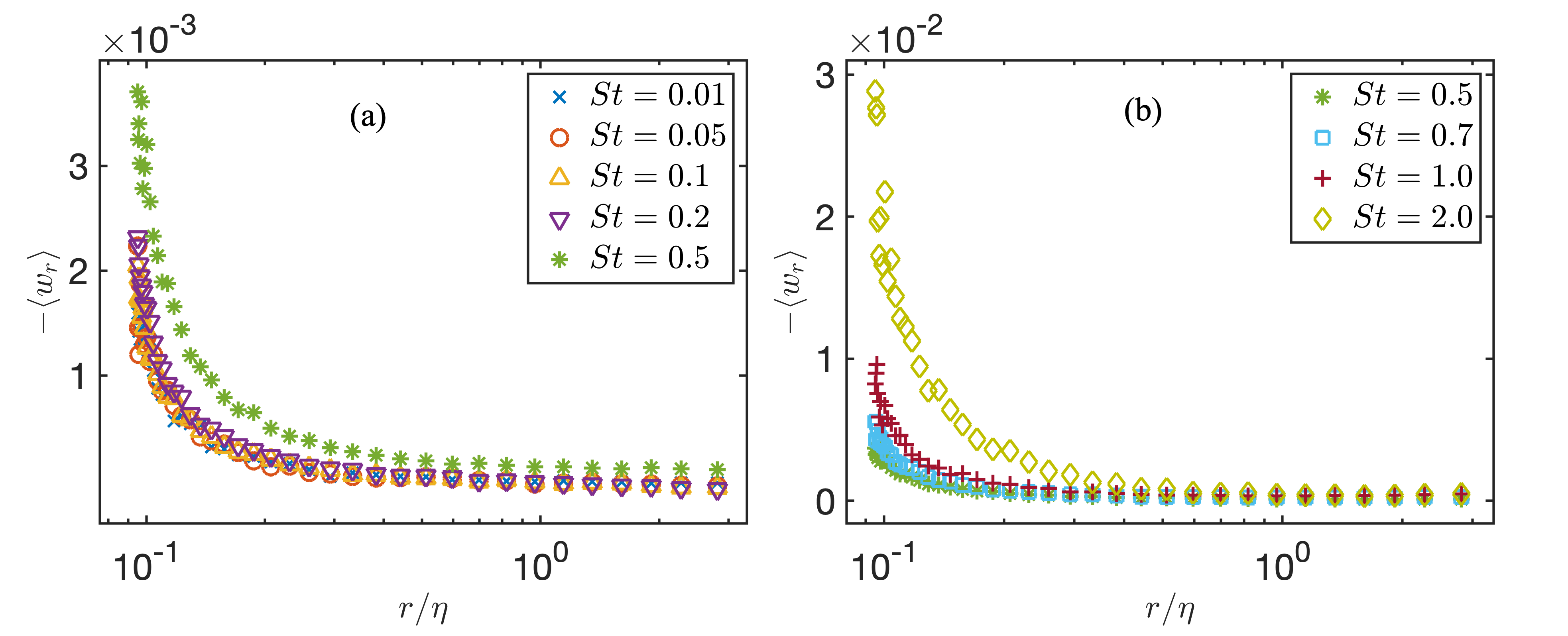}}
\caption{\xh{The MRVs for particles with different Stokes number.} In (a), $St=0.01$ to 0.5. In (b), $St=0.5$ to 2.0. $r$ is the separation distance of particles and $\eta$ is the Kolmogorov length scale of the turbulent flow.}
\label{Fig.mrv_st}
\end{figure}

In order to study the variation of the MRV with the particle Stokes number, particles with different $St$ are introduced into the system respectively, the range of $St$ is from 0.01 to 2.0. In each case, the turbulent Reynolds number and particle size are the same. The MRV for particles with different $St$ is shown in Figure \ref{Fig.mrv_st}.

\begin{figure}
	\centerline{
	\includegraphics[width=0.95\textwidth]{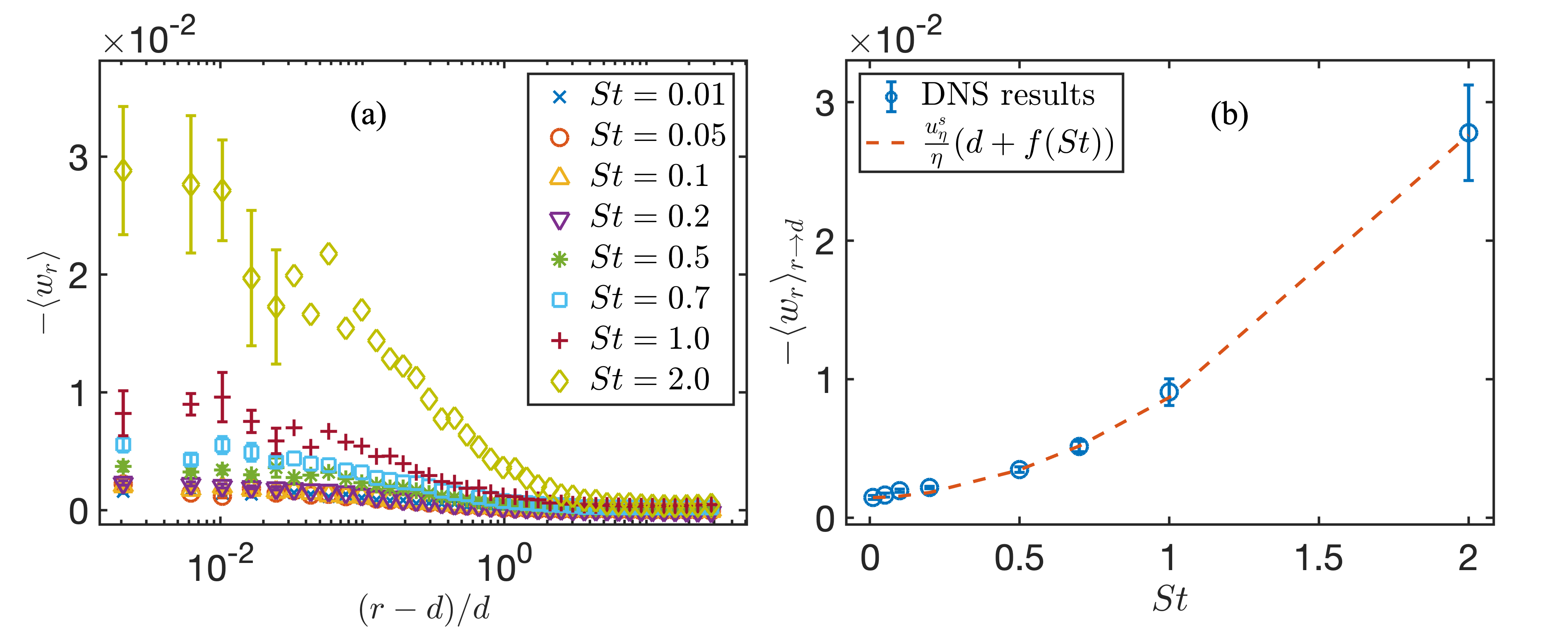}}
	\caption{(a) The MRV versus $(r-d)/d$ for particles with different Stokes number. $r$ is the particle separation distance and $d$ is the particle diameter. (b) The magnitude of MRV at $r=d$ for particles with different Stokes numbers. Blue $\circ$: the value of MRV from DNS results at $r-d\sim0.04d$. Red dashed line: prediction by Eq.\ref{Eq.wr_st}.}
	\label{Fig.Fvalue_st_all}
\end{figure}

First we can see from these figures that the magnitude of MRV increases gradually as the separation distance $r$ decreases, which is consistent with the results in \citep{saw2022intricate}. Furthermore, it also shows that for particles with $St\ll1$, the MRV barely changes with increasing Stokes number. However, when $St\geq0.2$, the magnitude of MRV increases significantly as $St$ increases. In order to illustrate the trend of the MRV when $r$ is close to $d$, the MRV against $(r-d)/d$ for particles with different Stokes number is shown in Figure \ref{Fig.Fvalue_st_all} (a). We can see that with the decrease of the particle separation distance $r$, the magnitude of MRV increases sharply at the first time. Then, it becomes flatten when $r$ is close to the particle diameter $d$. According to this phenomenon, \xh{we assume that the value of MRV at $r-d\approx0.04d$ can be regarded as the value of that at $r=d$, which we call $\langle w_r\rangle\vert_{r\to d}$.} The relationship between $\langle w_r\rangle\vert_{r\to d}$ and the particle Stokes number is shown in Figure \ref{Fig.Fvalue_st_all} (b), where the value of MRV is normalized by the Kolmogorov velocity $u_{\eta}$. We can see that $\langle w_r\rangle\vert_{r\to d}$ follows a convex function with $St$.

When the Stokes number is much less than \ew{unity}, the trajectory of inertial particles is close to \ew{that of fluid} particles and \ew{thus are strongly} influenced by the local flow velocity. In small scale range, the relative velocity of particles can be obtained by sampling the gradient of turbulent velocity\citep{ireland2016effect,biferale2014deformation}. \ew{Thus, in the limit of small Stokes number, MRV is nearly $St$-independent.} While for particles with larger $St$, they have larger \ew{kinetic} relaxation time, and the motion of particles at current time will be affected by the flow \ew{velocities encountered on their historical trajectories}, which is called path-history effect. In this situation, the correlation between \ew{velocities of nearby particles} is low and a large relative velocity is obtained at contact. This phenomenon is called sling effect or caustic \citep{falkovich2002acceleration,wilkinson2005caustics}. \ew{Earlier} studies show that when $St>0.2$, the sling effect is getting pronounced \citep{falkovich2007sling,ireland2016effect}, which is consistent with results in this paper.

Based on the above theoretical \ew{understanding}, we \ew{expect the trend of MRV versus Stokes number to be related to a "nominal" or "resonant" length scale which could be imagined as the nominal spatial scale (in the turbulence's energy spectrum) at which the inertial particles derive their momentum (or energy) from the advecting flow. This resonant scale is simply the particle separation $r$ when Stokes number is near zero due to passive advection, but it should increase with $St$ for heavier particles. 
We assume that for the Stokes numbers selected in this paper, this resonant scale does not exceed the Kolmogorov length scale $\eta$, so that the statistics of the flow velocity can be regarded as a linear function of spatial scale. Then, we explore a relationship between the value of MRV at $r=d$ with the particle Stokes number of the form:
}
\begin{equation}
	\langle w_r\rangle_{r=d}=\frac{u_{\eta}^s}{\eta}(d+\Delta l(St))
	\label{Eq.wr_st}
\end{equation}

where $u_{\eta}^s=\frac{u_{\eta}}{C\sqrt{15}}$, which comes from the second-order structure function of turbulent velocity \ew{at the small scale limit. We found $C=1.68$ from the DNS (it represents the difference between the average value and the root mean square). In Eq.\ref{Eq.wr_st}, $\Delta l$ represents the difference between the resonant scale at which particles couple most efficiently to the flow and $d$, and has the form $\Delta l=\alpha St^{\beta}$. By performing curve fitting on the DNS results (Fig.~\ref{Fig.Fvalue_st_all}b), we get $\alpha=4.43d$, $\beta=1.857$. Fig~.\ref{Fig.Fvalue_st_all}(b) shows the plot of Eq.\ref{Eq.wr_st} using the above obtained values of ($\alpha, \beta$) and the corresponding DNS result. Bringing $St=2.0$ into the function yields $\Delta l(St=2.0)=0.0146\sim\eta$, which satisfies the assumption made above.
}



\subsection{Reynolds number dependence}\label{sec.Re}

\begin{figure}
\centering
\includegraphics[width=0.95\textwidth]{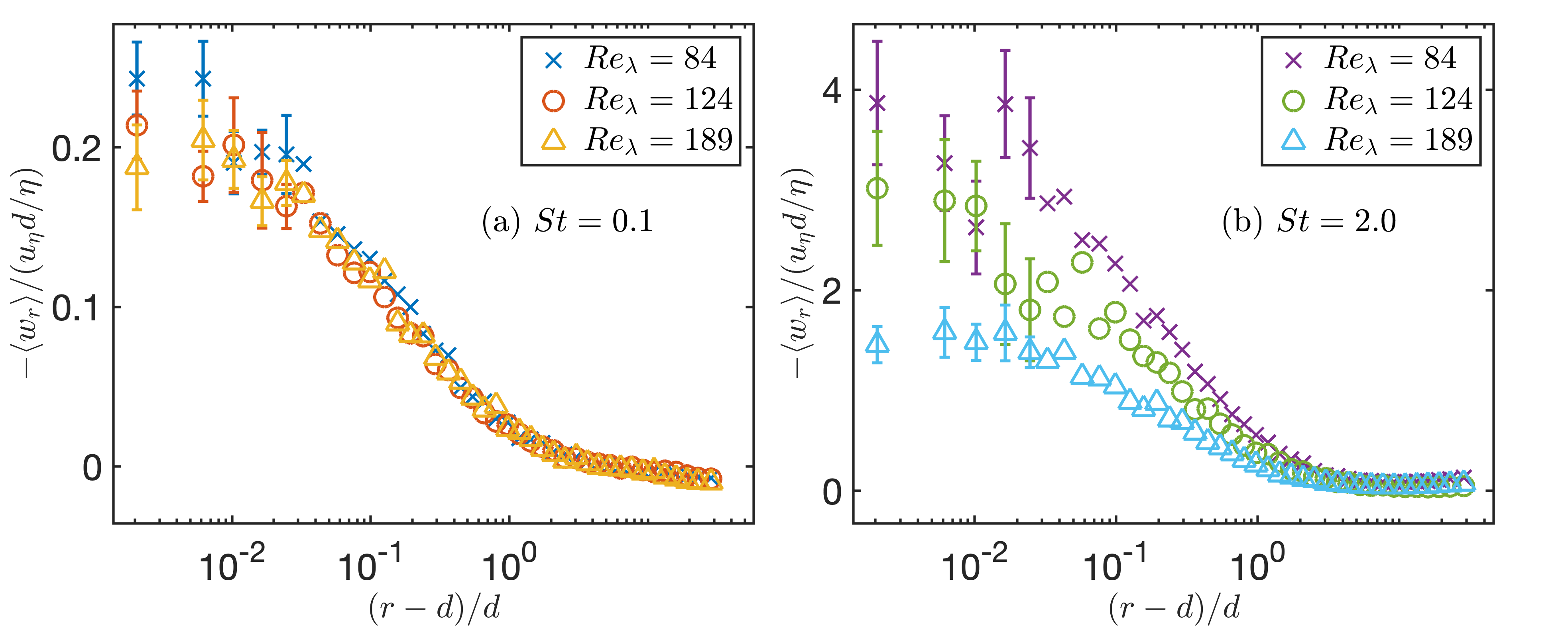}
\caption{The MRV versus $(r-d)/d$ for particles in different turbulent flow with $Re_{\lambda}=84$, 124, and 189 respectively. The value of MRV is normalized by $u_{\eta}(r/\eta)$. In (a), $St=0.1$. In (b), $St=2.0$.}
\label{Fig.mrv_Re}
\end{figure}

First, we study the MRV for particles with $St\ll1$ (e.g., $St=0.1$) in different $Re_{\lambda}$ cases, which is shown in Figure \ref{Fig.mrv_Re} (a). According to the results in Section \ref{sec.St}, particles with $St=0.1$ are mainly affected by the local turbulent velocity. Defining the characteristic local velocity of turbulent flow at $r=d$ as $u_{\eta}d/\eta$, the MRV in Figure \ref{Fig.mrv_Re} (a) is normalized by this value. It can be seen that the MRV in different $Re_{\lambda}$ cases collapse, which indicates that the MRV is independent to turbulent Reynolds number for $St\ll1$. While for larger $St$ (e.g. $St=2.0$), which is shown in Figure \ref{Fig.mrv_Re} (b), the magnitude of MRV, which is also normalized by $u_{\eta}d/\eta$, decreases with increasing $Re_{\lambda}$. Based on the theory in Eq.\ref{Eq.wr_st}, this result indicates that the spatial scale where particles are affected by the non-local flow in their historical trajectories decreases with $Re_{\lambda}$ increasing.

\begin{figure}
\centering
\subfigure[]{
\includegraphics[width=0.45\textwidth]{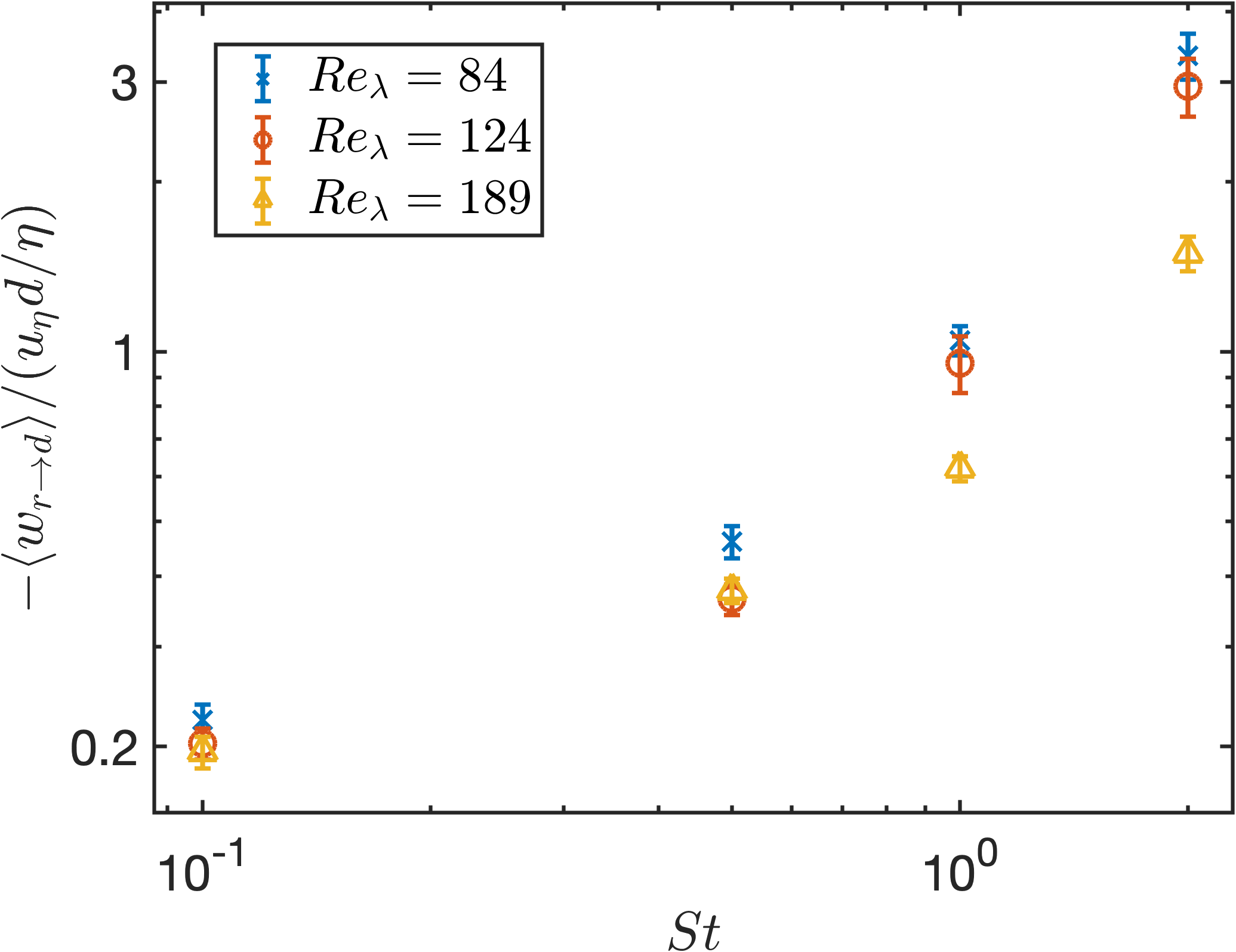}
\label{Fig.Fvalue_Re}}
\subfigure[]{
\includegraphics[width=0.45\textwidth]{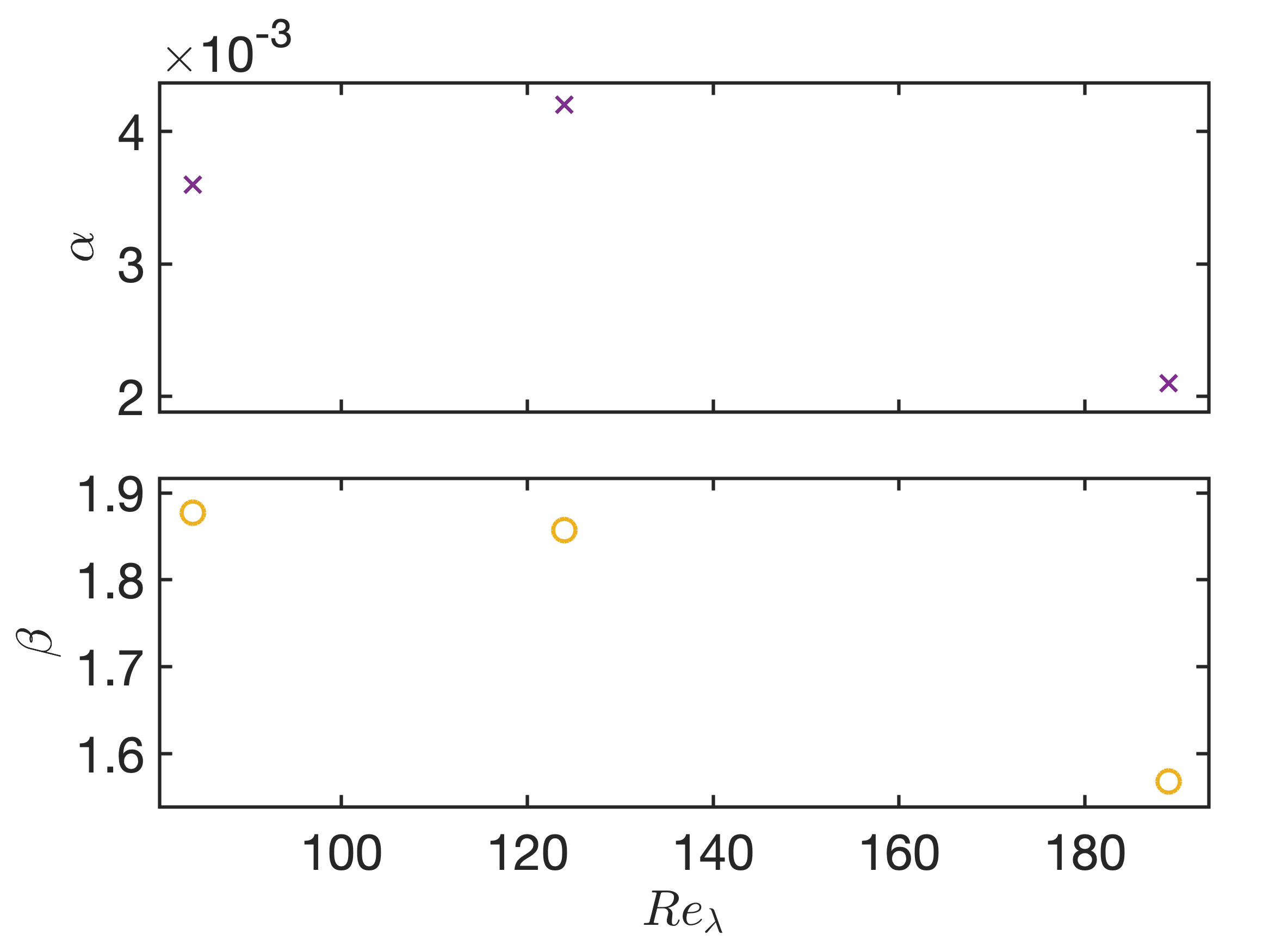}
\label{Fig.Fst_Re}}
\caption{(a). The relationship between $St$ and the \ew{magnitude} of MRV at $r-d\sim0.04d$ in different $Re_{\lambda}$ cases. The value of MRV is normalized by the characteristics local turbulent velocity $u_{\eta}d/\eta$. (b). The coefficients $\alpha$ and $\beta$ in $f(St)$ of Eq.\ref{Eq.wr_st} in different $Re_{\lambda}$ cases.}
\label{Fig.mrv_Re_value}
\end{figure}

In \citep{ireland2016effect}, they studied \ew{the average of} negative particle relative velocities $\langle w_r\vert w_r<0\rangle$ and found that \ew{its} values decrease with the increase of turbulent Reynolds number, but their analysis focused on particles with large $St$ (e.g., $St>3.0$), ignoring the particles results of smaller $St$. In order to be able to quantitatively analyze the influence of the turbulent Reynolds number, the value of MRV at $r\approx d$ for particles with $St=0.5$ and 1.0 in different $Re_{\lambda}$ cases are calculated and are compared with $St=0.1$ and 2.0 results, which is shown in Figure \ref{Fig.Fvalue_Re}. We can see that the magnitude of MRV increases with increasing $St$ in different $Re_{\lambda}$ cases. However, the growth rate of MRV with $St$ gradually slows down as $Re_{\lambda}$ increases. This result shows that for particles with larger Stokes number, the distance between the spatial scale affected by non-local flow and the particle collision radius $d$ is not only related to the particle Stokes number, but also is a function of turbulent Reynolds number. The coefficients in $f(St)$ are obtained by performing curve fitting on the DNS results in Figure \ref{Fig.Fvalue_Re}, which is shown in Figure \ref{Fig.Fst_Re}. $\beta$ decreases as the Reynolds number increases, while the value of $\alpha$ first increases and then decreases.

\subsection{Particle size dependence}\label{sec.Size}

\begin{figure}
	\centerline{
	\includegraphics[width=0.6\textwidth]{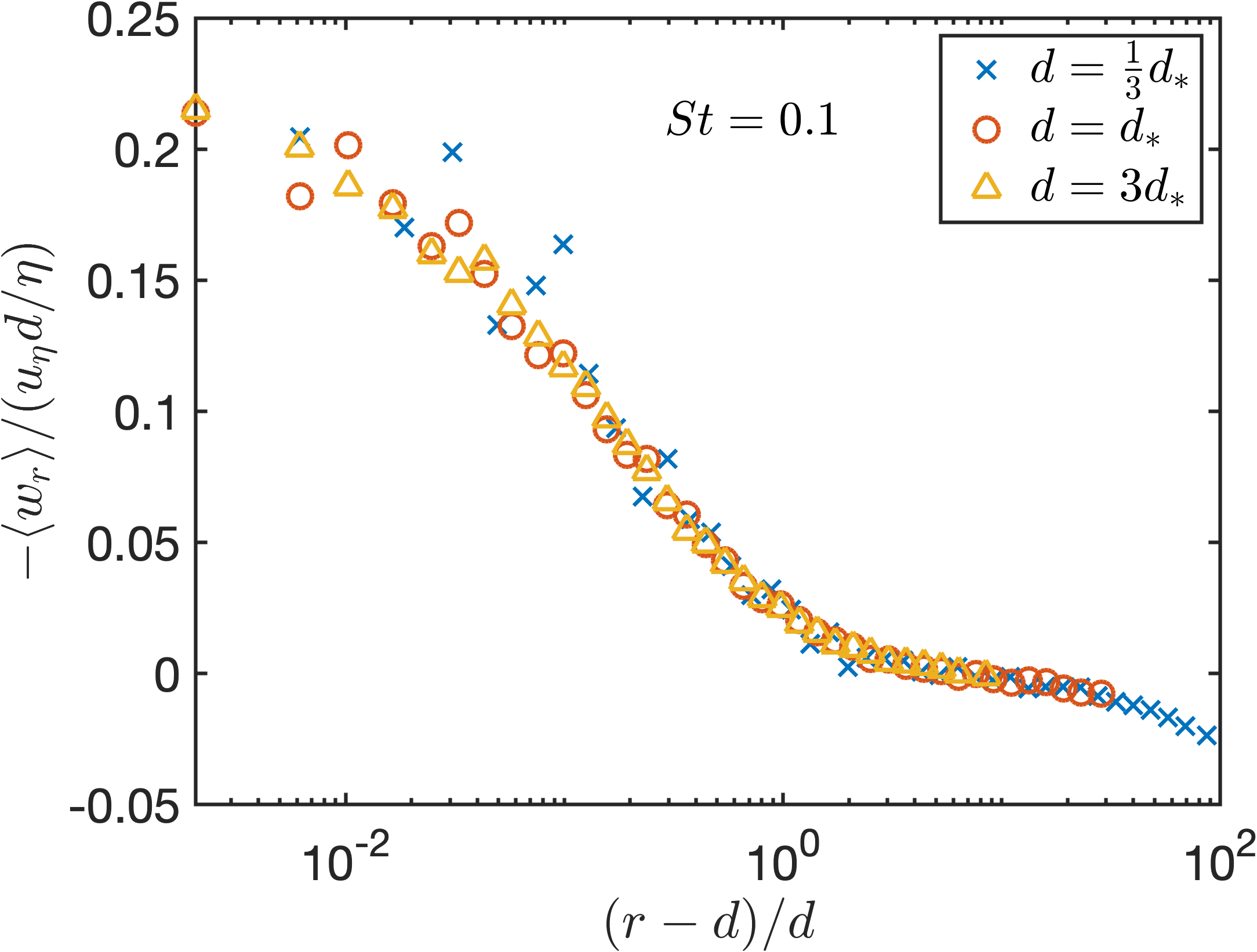}}
	\caption{The relationship between $(r-d)/d$ and the MRV for particles with different diameters. The value of the MRV is normalized by $u_{\eta}(r/\eta)$.}
	\label{Fig.mrv_size_d}
\end{figure}

The MRV for particles with different diameter is shown in Figure \ref{Fig.mrv_size_d}. The particle Stokes number is 0.1 and $Re_{\lambda}=124$ in three cases. The magnitude of MRV is normalized by the characteristic local turbulent velocity $u_{\eta}d/\eta$ and the MRV is plotted as a function of $(r-d)/d$. From this figure, we can see that the MRV for different diameter cases collapse. This result suggests that, despite the particle diameter is different, the particles with $St\ll1$ are still mainly affected by the local turbulent velocity. The variance of the MRV with the separation distance $r$ due to particle collision-coagulation is the same for different particle diameter.

\subsection{Extension of the MRV phenomenological model \ew{for finite $St$ particles}}\label{sec.Extension}

\begin{figure}
	\centerline{
	\includegraphics[width=0.95\textwidth]{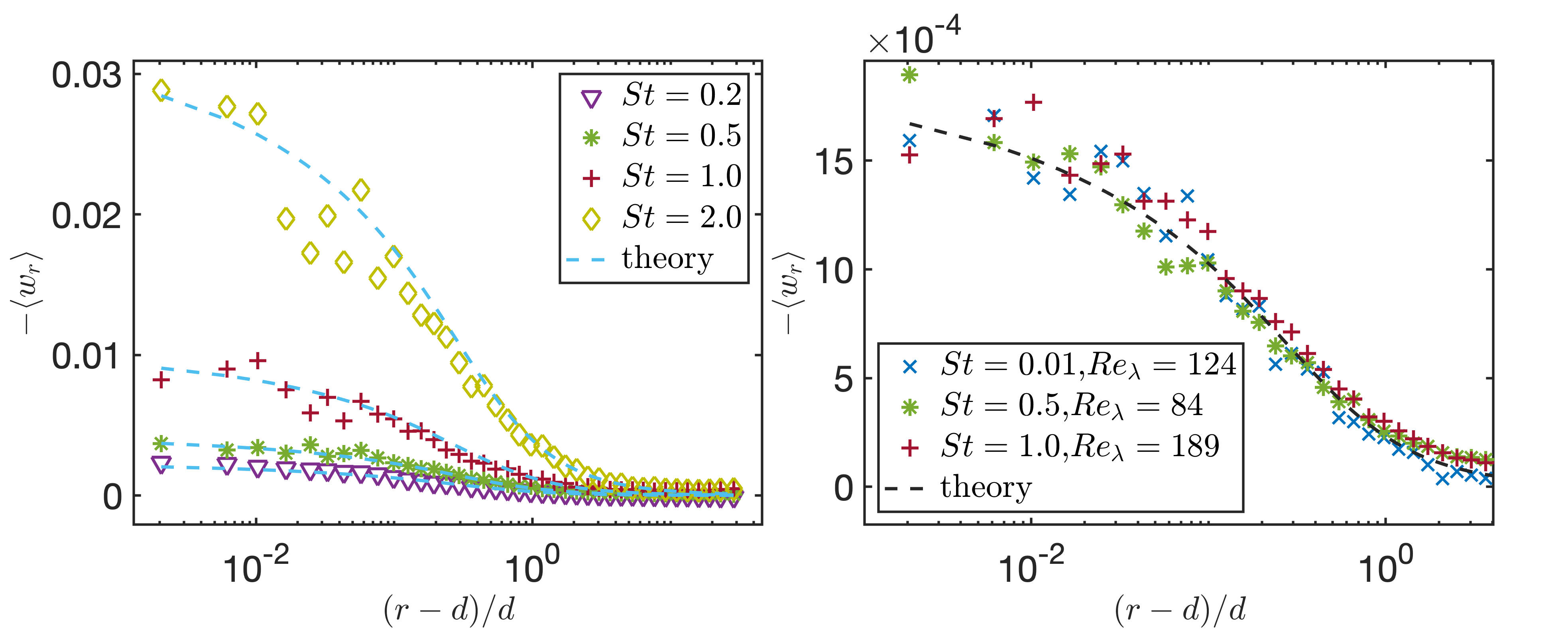}}
	\caption{The comparison between the theoretical model with DNS results. (a). Dashed lines: theoretical predictions via: $\gamma(St) \cdot \langle w_r\rangle^{theory}_{St \ll 0}$, where $\gamma (St)$ is a coefficient related to particle Stokes number derived from Eq.\ref{eq.gamma} and $\langle w_r\rangle^{theory}_{St \ll 0}$ is the "zero-Stokes" MRV phenomenological model Eq.\ref{Eq.wr_fit}. \ew{Colored symbols: DNS results for the MRV for particles with different $St$.} (b). The DNS-produced MRVs for different $St$ and $Re_{\lambda}$ cases are vertically translated to the position where $St=0.01$ case lies (color points). Dashed line: the prediction of MRV from Eq.\ref{Eq.wr_fit}, \ref{Eq.Ptheta} and \ref{Eq.pp}.}
	\label{Fig.model}
\end{figure}

\ew{The phenomenological model described in Section \ref{sec.Model} may be extended to predict the MRV in cases of larger (finite) Stokes numbers, if the model is multiplied by a pre-factor $\gamma (St)$, which
according to this Eq.~\ref{Eq.wr_st}, can be determined as:}
\begin{equation}
	\gamma=1+\frac{\Delta l(St)}{d}=1+\frac{\alpha}{d}St^{\beta}.
	\label{eq.gamma}
\end{equation}
Combining $\gamma$ with the MRV model (Eq.\ref{Eq.wr_fit}, \ref{Eq.Ptheta}, and \ref{Eq.pp}), \ew{we make quantitative predictions for the MRV for different Stokes numbers, and compare them with the DNS results in Figure \ref{Fig.model} (a). We see that the theoretical model is in good agreement with the DNS results. 
}

\ew{Since a definite relationship between $\Delta l$ and $Re_{\lambda}$ is unknown, in order to evaluate the model's ability to address possible $Re_{\lambda}$ effects, in Figure \ref{Fig.model} (b), the DNS produced MRV for the ($St=0.5$, $Re_{\lambda}=84$) case and ($St=1.0$, $Re_{\lambda}=189$) case are respectively translated vertically to the position where the ($St=0.01$, $Re_{\lambda}=124$) case lies. What is striking in this figure is that the MRV for different cases has the same shape, which is also predicted by the model in the range of $r<2d$. This suggests that the same "$\gamma \, \times\, model$" program is sufficient to also capture Reynolds number effects once the correct dependence of $\gamma$ on $R_{\lambda}$ is known.  
}

\section{Conclusion}

In this paper, the mean radial relative velocity (MRV) of inertial particles is investigated 
when the particle collision-coagulation is considered. First, the direct numerical simulation is used to investigate the relationship between the MRV with the particle Stokes number, the turbulent Reynolds number, and the particle diameter. The results show that the magnitude of MRV increases with increasing $St$, and the growth is pronounced when $St>0.2$. We attribute this increase to the scale at which particles with different inertia are affected by non-local turbulent flow in their historical trajectories. Based on this analysis, the relationship function of the value of MRV at $r=d$ with $St$ is obtained: $\langle w_r\rangle_{r=d}=\frac{u_{\eta}^s}{\eta}(d+f(St))$, where $f(St)$ represents the distance between the scale at which particles are affected by non-local turbulent velocity and their collision radius (e.g., particle diameter $d$). Next, we study the relationship between the value of MRV and the turbulent Reynolds number, and also analyze it from the perspective of the scale change of the non-local turbulent flow that affects inertial particles. We find that for particles whose Stokes number is much less than 1, they are mainly affected by the local turbulent velocity, and $f(St)$ is independent of $Re_{\lambda}$. While for particles with larger $St$, the coefficients $\alpha$ and $\beta$ in $f(St)$ decrease as $Re_{\lambda}$ increases. However, since this paper only selects three cases with different $Re_{\lambda}$, it is impossible to obtain an accurate mathematical relationship between $f(St)$ and $Re_{\lambda}$, which needs further research in the future. In addition, this paper analyzes the MRV for particles with different diameters $d$. The result shows that for particles with the same $St$, although the diameters are different, the MRV under the same $d/r$ has the same value, and $f(St)$ is independent of the particle diameter.

\ew{Finally, we make a correction (improvement) on the MRV phenomenological model proposed by \cite{saw2022intricate} to captures the effect of collision-coagulation on MRV in the limit of $St \ll 1$. As a result, 
the corrected model, coupled with proper inputs from the turbulent flow's velocity statistics, makes prediction that is in close agreement with the DNS results, an outcome that is substantially better than that in \citep{saw2022intricate}. Furthermore, based on the relationship between the value of MRV at $r=d$ and $St$ mentioned above, a coefficient $\gamma$ is defined. We found that by incorporating $\gamma$ into the MRV model, the MRV for cases with any $St$ can be accurately predicted (up to $St=2$) . Our results also suggest that the model could also accurately account for
any possible Reynolds number effect through a simple generalization.}

\backsection[Acknowledgements]{This work was mainly supported by the National Natural Science Foundation of China (grand 11872382) and by the Thoudand Young Program of China.}

\bibliography{mrv}

\begin{thebibliography}{24}
\expandafter\ifx\csname natexlab\endcsname\relax\def\natexlab#1{#1}\fi
\def\au#1{#1} \def\ed#1{#1} \def\yr#1{#1}\def\at#1{#1}\def\jt#1{\textit{#1}}
  \def\bt#1{#1}\def\bvol#1{\textbf{#1}} \def\vol#1{#1} \def\pg#1{#1}
  \def\publ#1{#1}\def\arxiv#1{#1}\def\org#1{#1}\def\st#1{\textit{#1}}

\bibitem[Bec {\em et~al.\/}(2010)Bec, Biferale, Cencini, Lanotte \&
  Toschi]{bec2010intermittency}
{\sc \au{Bec, J}, \au{Biferale, L}, \au{Cencini, M}, \au{Lanotte, AS} \&
  \au{Toschi, F}} \yr{2010}  \at{Intermittency in the velocity distribution of
  heavy particles in turbulence}.  \jt{Journal of Fluid Mechanics}  \bvol{646},
   \pg{527--536}.

\bibitem[Bewley {\em et~al.\/}(2013)Bewley, Saw \&
  Bodenschatz]{bewley2013observation}
{\sc \au{Bewley, Gregory~P}, \au{Saw, Ewe-Wei} \& \au{Bodenschatz, Eberhard}}
  \yr{2013}  \at{Observation of the sling effect}.  \jt{New Journal of Physics}
   \bvol{15}~(8),  \pg{083051}.

\bibitem[Biferale {\em et~al.\/}(2014)Biferale, Meneveau \&
  Verzicco]{biferale2014deformation}
{\sc \au{Biferale, Luca}, \au{Meneveau, Charles} \& \au{Verzicco, Roberto}}
  \yr{2014}  \at{Deformation statistics of sub-kolmogorov-scale ellipsoidal
  neutrally buoyant drops in isotropic turbulence}.  \jt{Journal of fluid
  mechanics}  \bvol{754},  \pg{184--207}.

\bibitem[Bragg \& Collins(2014)]{bragg2014new}
{\sc \au{Bragg, Andrew~D} \& \au{Collins, Lance~R}} \yr{2014}  \at{New insights
  from comparing statistical theories for inertial particles in turbulence: Ii.
  relative velocities}.  \jt{New Journal of Physics}  \bvol{16}~(5),
  \pg{055014}.

\bibitem[Eswaran \& Pope(1988)]{eswaran1988examination}
{\sc \au{Eswaran, Vinayak} \& \au{Pope, Stephen~B}} \yr{1988}  \at{An
  examination of forcing in direct numerical simulations of turbulence}.
  \jt{Computers \& Fluids}  \bvol{16}~(3),  \pg{257--278}.

\bibitem[Falkovich {\em et~al.\/}(2002)Falkovich, Fouxon \&
  Stepanov]{falkovich2002acceleration}
{\sc \au{Falkovich, Gregory}, \au{Fouxon, A} \& \au{Stepanov, MG}} \yr{2002}
  \at{Acceleration of rain initiation by cloud turbulence}.  \jt{Nature}
  \bvol{419}~(6903),  \pg{151--154}.

\bibitem[Falkovich \& Pumir(2007)]{falkovich2007sling}
{\sc \au{Falkovich, Gregory} \& \au{Pumir, Alain}} \yr{2007}  \at{Sling effect
  in collisions of water droplets in turbulent clouds}.  \jt{Journal of the
  Atmospheric Sciences}  \bvol{64}~(12),  \pg{4497--4505}.

\bibitem[Grabowski \& Wang(2013)]{grabowski2013growth}
{\sc \au{Grabowski, Wojciech~W} \& \au{Wang, Lian-Ping}} \yr{2013}  \at{Growth
  of cloud droplets in a turbulent environment}.  \jt{Annual review of fluid
  mechanics}  \bvol{45},  \pg{293--324}.

\bibitem[Ireland {\em et~al.\/}(2016)Ireland, Bragg \&
  Collins]{ireland2016effect}
{\sc \au{Ireland, Peter~J}, \au{Bragg, Andrew~D} \& \au{Collins, Lance~R}}
  \yr{2016}  \at{The effect of reynolds number on inertial particle dynamics in
  isotropic turbulence. part 1. simulations without gravitational effects}.
  \jt{Journal of Fluid Mechanics}  \bvol{796},  \pg{617--658}.

\bibitem[Johansen {\em et~al.\/}(2009)Johansen, Youdin \&
  Mac~Low]{johansen2009particle}
{\sc \au{Johansen, Anders}, \au{Youdin, Andrew} \& \au{Mac~Low, Mordecai-Mark}}
  \yr{2009}  \at{Particle clumping and planetesimal formation depend strongly
  on metallicity}.  \jt{The Astrophysical Journal}  \bvol{704}~(2),  \pg{L75}.

\bibitem[Maxey \& Riley(1983)]{maxey1983equation}
{\sc \au{Maxey, Martin~R} \& \au{Riley, James~J}} \yr{1983}  \at{Equation of
  motion for a small rigid sphere in a nonuniform flow}.  \jt{The Physics of
  Fluids}  \bvol{26}~(4),  \pg{883--889}.

\bibitem[Meng \& Saw(2023)]{meng2023sharp}
{\sc \au{Meng, Xiaohui} \& \au{Saw, Ewe-Wei}} \yr{2023}  \at{Sharp depletion of
  radial distribution function of particles due to collision and coagulation
  inside turbulent flow: A systematic study}.  \jt{Physical Review Fluids}
  \bvol{8}~(8),  \pg{084304}.

\bibitem[Pan {\em et~al.\/}(2011)Pan, Padoan, Scalo, Kritsuk \&
  Norman]{pan2011turbulent}
{\sc \au{Pan, Liubin}, \au{Padoan, Paolo}, \au{Scalo, John}, \au{Kritsuk,
  Alexei~G} \& \au{Norman, Michael~L}} \yr{2011}  \at{Turbulent clustering of
  protoplanetary dust and planetesimal formation}.  \jt{The Astrophysical
  Journal}  \bvol{740}~(1),  \pg{6}.

\bibitem[Rogallo(1981)]{rogallo1981numerical}
{\sc \au{Rogallo, Robert~Sugden}} \yr{1981} {\em Numerical experiments in
  homogeneous turbulence\/}, ,  \vol{vol. 81315}.  \publ{National Aeronautics
  and Space Administration}.

\bibitem[Rosa {\em et~al.\/}(2013)Rosa, Parishani, Ayala, Grabowski \&
  Wang]{rosa2013kinematic}
{\sc \au{Rosa, Bogdan}, \au{Parishani, Hossein}, \au{Ayala, Orlando},
  \au{Grabowski, Wojciech~W} \& \au{Wang, Lian-Ping}} \yr{2013}  \at{Kinematic
  and dynamic collision statistics of cloud droplets from high-resolution
  simulations}.  \jt{New Journal of Physics}  \bvol{15}~(4),  \pg{045032}.

\bibitem[Saffman \& Turner(1956)]{saffman1956collision}
{\sc \au{Saffman, PGF} \& \au{Turner, JS}} \yr{1956}  \at{On the collision of
  drops in turbulent clouds}.  \jt{Journal of Fluid Mechanics}  \bvol{1}~(1),
  \pg{16--30}.

\bibitem[Saw \& Meng(2022)]{saw2022intricate}
{\sc \au{Saw, Ewe-Wei} \& \au{Meng, Xiaohui}} \yr{2022}  \at{Intricate
  relations among particle collision, relative motion and clustering in
  turbulent clouds: computational observation and theory}.  \jt{Atmospheric
  Chemistry and Physics}  \bvol{22}~(6),  \pg{3779--3788}.

\bibitem[Saw {\em et~al.\/}(2008)Saw, Shaw, Ayyalasomayajula, Chuang \&
  Gylfason]{saw2008inertial}
{\sc \au{Saw, Ewe~Wei}, \au{Shaw, Raymond~A}, \au{Ayyalasomayajula,
  Sathyanarayana}, \au{Chuang, Patrick~Y} \& \au{Gylfason, Armann}} \yr{2008}
  \at{Inertial clustering of particles in high-reynolds-number turbulence}.
  \jt{Physical review letters}  \bvol{100}~(21),  \pg{214501}.

\bibitem[Shaw(2003)]{shaw2003particle}
{\sc \au{Shaw, Raymond~A}} \yr{2003}  \at{Particle-turbulence interactions in
  atmospheric clouds}.  \jt{Annual Review of Fluid Mechanics}  \bvol{35}~(1),
  \pg{183--227}.

\bibitem[Smith {\em et~al.\/}(2002)Smith, Nathan, Zhang \&
  Nobes]{smith2002significance}
{\sc \au{Smith, NL}, \au{Nathan, GJ}, \au{Zhang, DK} \& \au{Nobes, DS}}
  \yr{2002}  \at{The significance of particle clustering in pulverized coal
  flames}.  \jt{Proceedings of the Combustion Institute}  \bvol{29}~(1),
  \pg{797--804}.

\bibitem[Squires \& Eaton(1991)]{squires1991preferential}
{\sc \au{Squires, Kyle~D} \& \au{Eaton, John~K}} \yr{1991}  \at{Preferential
  concentration of particles by turbulence}.  \jt{Physics of Fluids A: Fluid
  Dynamics}  \bvol{3}~(5),  \pg{1169--1178}.

\bibitem[Sundaram \& Collins(1997)]{sundaram1997collision}
{\sc \au{Sundaram, Shivshankar} \& \au{Collins, Lance~R}} \yr{1997}
  \at{Collision statistics in an isotropic particle-laden turbulent suspension.
  part 1. direct numerical simulations}.  \jt{Journal of Fluid Mechanics}
  \bvol{335},  \pg{75--109}.

\bibitem[Wilkinson \& Mehlig(2005)]{wilkinson2005caustics}
{\sc \au{Wilkinson, M} \& \au{Mehlig, Bernhard}} \yr{2005}  \at{Caustics in
  turbulent aerosols}.  \jt{Europhysics Letters}  \bvol{71}~(2),  \pg{186}.

\bibitem[Wilkinson {\em et~al.\/}(2006)Wilkinson, Mehlig \&
  Bezuglyy]{wilkinson2006caustic}
{\sc \au{Wilkinson, Michael}, \au{Mehlig, Bernhard} \& \au{Bezuglyy, Vlad}}
  \yr{2006}  \at{Caustic activation of rain showers}.  \jt{Physical review
  letters}  \bvol{97}~(4),  \pg{048501}.

\end{thebibliography}
\bibliographystyle{jfm}

\end{document}